\documentclass[aps,prb,10pt,twocolumn,showpacs,preprintnumbers,amsmath,amssymb,superscriptaddress,floatfix]{revtex4-2}

\usepackage{amsmath}    
\usepackage{amssymb,environ}
\usepackage{graphicx}   
\usepackage{verbatim}   
\usepackage{color}      

\usepackage[normalem]{ulem}
\usepackage{natbib}
\usepackage{enumitem}
\usepackage{multirow}

\usepackage{epsfig}
\usepackage{textcomp}
\usepackage{wasysym}
\usepackage{array}
\usepackage{color}

\usepackage{hyperref}   

\begin{document}

\newcommand{\ev}[0]{\mathbf{e}}
\newcommand{\cv}[0]{\mathbf{c}}
\newcommand{\fv}[0]{\mathbf{f}}
\newcommand{\Rv}[0]{\mathbf{R}}
\newcommand{\Tr}[0]{\mathrm{Tr}}
\newcommand{\ud}[0]{\uparrow\downarrow}
\newcommand{\du}[0]{\downarrow\uparrow}
\newcommand{\Uv}[0]{\mathbf{U}}
\newcommand{\Iv}[0]{\mathbf{I}}
\newcommand{\Hv}[0]{\mathbf{H}}
\newcommand{\kv}[0]{\mathbf{k}}
\newcommand{\qv}[0]{\mathbf{q}}
\newcommand{\uu}[1]{\underline{\underline{#1}}}

\setlength{\jot}{2mm}

\newcommand{\jav}[1]{#1}
\newcommand{\ds}[1]{{\color{blue}#1}}

\title{Systematic Schrieffer-Wolff-transformation approach to Josephson junctions: quasiparticle effects and Josephson harmonics}

\author{\'Ad\'am B\'acsi}
\email{bacsi.adam@sze.hu}
\affiliation{Department of Mathematics and Physics, Sz\'echenyi Istv\'an University, 9026 Gy\H or, Hungary}
\affiliation{MTA-BME Lendület "Momentum" Open Quantum Systems Research Group, Institute of Physics, Budapest University of Technology and Economics, Műegyetem rkp. 3., H-1111, Budapest, Hungary}

\author{Teodor Ili\v cin}
\affiliation{Jo\v zef Stefan Institute, Jamova 39, SI-1000 Ljubljana, Slovenia}
\affiliation{Faculty of Mathematics and Physics, University of Ljubljana, Jadranska 19, SI-1000 Ljubljana, Slovenia}

\author{Rok \v Zitko}
\affiliation{Jo\v zef Stefan Institute, Jamova 39, SI-1000 Ljubljana, Slovenia}

\begin{abstract}
We use the Schrieffer-Wolff transformation (SWT) to analyze Josephson junctions between superconducting leads described by the charge-conserving BCS theory. Starting from the single-electron tunneling terms, we directly recover the conventional effective Hamiltonian, $-E_J\cos\hat{\varphi}$, with an operator-valued phase bias $\hat{\varphi}$. The SWT approach has the advantage that it can be systematically extended to more complex scenarios. We show that if a Bogoliubov quasiparticle is present its motion couples to that of Cooper pairs, introducing correlated dynamics that reshape the energy spectrum of the junction. Furthermore, higher-order terms in the SWT  naturally describe Josephson harmonics, whose amplitudes are directly related to the microscopic properties of the superconducting leads and the junction. We derive expressions that could facilitate tuning the ratio between the different harmonics in a controlled way.
\end{abstract}

\maketitle

\section{Introduction}
Josephson junctions (JJs) are key components of superconducting quantum devices \cite{krantz2019,kjaergaard2020,soloviev2021,Aguado2024}, as the non-linear characteristics arising from Cooper pair (CP) tunneling play a crucial role in establishing a well-separated excitation energy between the ground state and the first excited state of the system.
In ideal operation, JJs rely solely on the coherent tunneling of CPs \cite{josephson1962,blais2021}. Under certain conditions, however, the pairs break into quasiparticles (QPs), the elementary excitations of superconductors, that disrupt the coherent behavior by tunneling across the junction. This phenomenon is known as the QP poisoning \cite{Catelani2012,Catelani_2022,Hays2024} and it is particularly important because the population of QPs is often much higher than expected from thermal equilibrium \cite{bespalov2016,glazman2021,catelani2024}.

Beside the experimental challenges of mitigating the QP poisoning \cite{Cardani2021, Pan2022, kamenov2024, connolly2024, Papic2024}, a range of theoretical questions concerning the origin of QPs, their dynamics, and their impact on device performance are still open \cite{Aumentado2023,PRXQuantumDevoret,Wang2014, houzet2019}. The central problem is how, given that the QPs are inevitably present in the system, they can be described in the effective models of transmons, flux qubits, and other JJ-based superconducting quantum devices.

One goal of this paper is to extend the conventional model of the JJs, $H=-E_J\cos\varphi$, to account for the presence of a QP. To achieve this, we employ the Schrieffer-Wolff transformation (SWT), a widely used perturbative method in theoretical physics \cite{SWSIAM1966,tperU1988,landi2024,blais2021,kenawy2022}. The SWT provides a systematic way to derive low-energy effective models through a unitary transformation followed by a projection onto the relevant low-energy subspace \cite{bravyi2011,zhang2022}. \jav{The novelty of our approach lies in the fact that the SWT allows the inclusion of QPs directly at the level of the effective Hamiltonian and describes the correlations between QPs and CPs in a quantum-coherent way, in contrast to previous studies based on master-equation methods \cite{Catelani2012}. Furthermore, the SWT approach highlights the underlying unitary transformation between the microscopic BCS theory and the resulting JJ Hamiltonian for the phase difference variable, and it directly provides operator expressions. This is in contrast to conventional perturbation theory methods which result in corrections to the energies and wavefunctions, from which the effective Hamiltonian (when needed) needs to be reconstructed, which is challenging in all but the simplest cases.} In the context of JJs, the SWT has been previously used, for example, to describe optomechanical coupling \cite{heikkila2014}, but not to derive the effective JJ Hamiltonian from the microscopic single-electron tunneling terms.

Here, we perform a SWT on the charge-conserving Bardeen-Cooper-Schrieffer (BCS) theory \cite{tinkhamPRB,josephson1962} and derive the low-energy model of the JJ. By initially restricting the low-energy subspace to QP-free states and performing a second-order transformation, we recover the conventional model $H=-E_J\cos\varphi$. 
To incorporate QPs, in Sec. \ref{sec:1qp} we relax this restriction by including states containing QPs in the low-energy subspace and derive the corresponding effective model for a junction populated by a single QP.
The SWT further provides a systematic framework to extend the theory to higher-order perturbations. In Sec. \ref{sec:4order}, we restrict again to the QP-free states in the low-energy subspace and perform the SWT to fourth order, enabling the description of Josephson harmonics \cite{secondharmonic2007,Willsch2024} and providing exact expressions for calculating effective-model parameters from microscopic quantities.

\section{Charge-conserving BCS theory}
\label{sec:model}
The microscopic theory of superconductivity starts with the attractive Hamiltonian
\begin{equation}
\begin{split}
H -\mu \hat{N} &= \sum_{k}\xi_k\left( c_{k\uparrow}^+ c_{k\uparrow} + c_{-k\downarrow}^+ c_{-k\downarrow}\right) \\ 
&- \frac{g}{N}\sum_{kk'}c_{k\uparrow}^+ c_{-k\downarrow}^+ c_{-k'\downarrow} c_{k'\uparrow}
\end{split}
\label{eq:ham4}
\end{equation}
where $\xi_k =\varepsilon_k - \mu$ is the non-interacting electron energy spectrum with $\mu$ the chemical potential, while $g$ characterizes the strength of the attractive interaction. In the BCS theory, the attractive interaction term is decoupled via a mean-field approximation using the order parameter $\frac{g}{N} \sum_{k} \langle c_{-k\downarrow} c_{k\uparrow} \rangle$. This leads to a model which does not commute with the total number of electrons $N_e = \sum_{k\sigma} c_{k\sigma}^+ c_{k\sigma}$ and therefore does not conserve charge. 

A slightly modified version of the mean-field approximation defines the order parameter as
\begin{equation}
\Delta = \frac{g}{N}\sum_{k} \langle S^+ c_{-k\downarrow} c_{k\uparrow}  \rangle
\end{equation}
where the operator $S^+$ creates a CP in the condensate \cite{tinkhamPRB,josephson1962}. The $S$ operator commutes with the fermionic operators and acts on an auxiliary Hilbert space of CPs which is distinct from the fermionic Hilbert space of QPs. In this theory, the states of the superconducting system have the tensor-product structure of $|\mbox{fermions}\rangle\otimes|M\rangle_c$ where $|M\rangle_c$ denotes the state of the condensate consisting of $M$ Cooper pairs. We note that because of $S^+S = I - |0\rangle_c\langle 0|_c$ where $|0\rangle_c$ is the state with no CP in the condensate, the operator $S$ is \textit{almost} unitary. Since the system is populated with a macroscopic number of CPs, the term $|0\rangle_c\langle 0|_c$ is negligible. Furthermore, the $S$ operator becomes a true unitary operator if we allow $M$ to be negative which we adopt here due to computational advantages. 

After taking the mean-field approximation, as described in the Appendix of Ref. \cite{praPofE}, the Bogoliubov transformation
\begin{equation}
d_{k\sigma} = u_k c_{k\sigma} - \sigma v_k S c_{-k\bar{\sigma}}^+
\label{eq:dop}
\end{equation}
leads to the diagonal form of the Hamiltonian
\begin{equation}
H - \mu \hat{N}=E_{GS} + \sum_{k\sigma}E_k d_{k\sigma}^+ d_{k\sigma}
\end{equation}
where $E_k = \sqrt{\xi_k^2 + |\Delta|^2}$ is the spectrum of the superconducting QPs. The ground-state energy $E_{GS}$ will henceforth be omitted as it does not affect the low-energy physics of the system. In Eq. \eqref{eq:dop}, $u_k = \sqrt{(1+\xi_k/E_k)/2}$ and $v_k = e^{i\delta}\sqrt{(1-\xi_k/E_k)/2}$ are the conventional Bogoliubov coefficients that diagonalize the mean-field Hamiltonian. Note that $d_{k\sigma}$ reduces the total charge of the system by one unit since its second term describes the annihilation of a CP accompanied by the creation of an electron. The total number of particles in the system is described by the operator $\hat{N} = \sum_{k\sigma}c_{k\sigma}^+c_{k\sigma} + 2\sum_M M |M\rangle_c\langle M|_c$, which commutes with the Hamiltonian, leading to the notion of charge-conserving BCS theory. Since the charge is conserved, the phase of the gap, $\delta$ which is at the same time the phase of $v_k$, can be freely chosen. 

The ground state of the mean-field Hamiltonian is given by
\begin{equation}
|\mathrm{GS},M\rangle = \prod_{k}\left(u_k + v_k S c_{k\uparrow}^+ c_{-k\downarrow}^+\right) |0\rangle \otimes |M\rangle_c
\label{eq:gs}
\end{equation}
where each term corresponds to a state with a total of $2M$ particles.

\section{Schrieffer-Wolff transformation in Josephson junctions}
\label{sec:SWbase}
A JJ consists of two superconducting leads separated by a thin insulating layer across which quantum tunneling can occur. The leads are modeled using the charge-conserving BCS Hamiltonians 
\begin{equation}
\begin{split}
H_L & = \sum_{k\sigma}E_k d_{k\sigma}^+ d_{k\sigma} + \mu_L \hat{N}_L \\
H_R & = \sum_{q\sigma}E_q d_{q\sigma}^+ d_{q\sigma} + \mu_R \hat{N}_R
\end{split}
\end{equation}
where we follow the conventional notation of the wavenumber $k$ in the left lead and $q$ in the right lead.
In the absence of QPs, the basis states are given by $|M_L,M_R\rangle = |\mathrm{GS},M_L\rangle\otimes |\mathrm{GS},M_R\rangle$ describing the ground state with $2M_L$ ($2M_R$) particles in the left (right) lead. These states constitute the low-energy subspace for the SWT. 

The perturbation is the tunneling Hamiltonian describing single-electron tunneling, which we express in terms of the Bogoliubov QP operators $d$ as
\begin{widetext}
\begin{equation}
\begin{split}
V = \sum_{kq\sigma}\left(T_{kq} c_{q\sigma}^+ c_{k\sigma} + h.c.\right)
=& \sum_{kq\sigma}\Big[T_{kq}\Big( u_k u_q d_{q\sigma}^+ d_{k\sigma} - v_q^* v_k S_R^+ S_L d_{k\sigma}^+ d_{q\sigma}  + \sigma v_q^* u_k \underbrace{S_R^+ d_{q\bar{\sigma}}d_{k\sigma}}_{\fbox{$\rightarrow A$}} + \sigma u_q v_k \underbrace{S_L d_{q\sigma}^+ d_{k\bar{\sigma}}^+}_{\fbox{$\rightarrow C$}}\Big) + \\
+& T_{kq}^*\Big( u_k u_q d_{k\sigma}^+ d_{q\sigma} - v_q v_k^* S_L^+ S_R d_{q\sigma}^+ d_{k\sigma} + \sigma v_k^* u_q \underbrace{S_L^+ d_{k\bar{\sigma}}d_{q\sigma}}_{\fbox{$\leftarrow A$}} + \sigma u_k v_q \underbrace{S_R d_{k\sigma}^+ d_{q\bar{\sigma}}^+}_{\fbox{$\leftarrow C$}}\Big)\Big].
\end{split}
\label{eq:Ht}
\end{equation}
\end{widetext}
The tunneling can be decomposed into eight different QP processes. The labeled processes in Eq. \eqref{eq:Ht} are those that affect the charge balance between the fermionic and condensate sectors, as they describe the (C)reation or (A)nnihilation of a pair of QPs and the opposite process for a CP in the condensate. For example, the term $\fbox{$\rightarrow A$}$ describes the annihilation of two quasiparticles, one from each lead, and simultaneous creation of a CP in the condensate of the right lead. The arrow indicates the direction of the overall charge transfer, here one charge from left to right lead.

In the SWT, we treat $V$ as a perturbation and perform a unitary rotation followed by a projection onto the low-energy subspace \cite{bravyi2011,zhang2022}. The rotation is applied to the total Hamiltonian $H=H_0 + V$, with $H_0 = H_L + H_R$, as
\begin{equation}
H_\mathrm{eff}=P_0 e^R H e^{-R} P_0
\label{eq:SW}
\end{equation}
where $P_0$ describes the projection onto the low-energy subspace of QP-free states. Assuming the perturbation $V$ and the rotation operator $R$ to be small, we expand Eq. \eqref{eq:SW} to second order in $V$:
\begin{equation}
H_\mathrm{eff}\approx H_0 + V + \left[R,H_0\right] + \left[R,V\right] + \frac{1}{2}\left[R,\left[R,H_0\right]\right]
\label{eq:expansion2}
\end{equation}
We follow the standard procedure \cite{Phillips_2012,bravyi2011} and choose $R$ to obey $[R,H_0] + V = 0$ ensuring that the first-order terms in the expansion vanish. The effective Hamiltonian then contains only second-order corrections. Its matrix elements in the low-energy subspace are given by
\begin{equation}
\begin{split}
& \langle n|H_\mathrm{eff}|m\rangle =  E_n\delta_{nm} + \\ &+ \frac{1}{2}\sum_{l}\langle n|V|l\rangle
\langle l|V|m\rangle \left(\frac{1}{E_n-E_l} + \frac{1}{E_m-E_l}\right)
\end{split}
\label{eq:Heff}
\end{equation}
where $E_n$ is the eigenvalue of the unperturbed Hamiltonian $H_0$. The states $|n\rangle$ and $|m\rangle$ are low-energy states, such as $|M_L,M_R\rangle$ with the energy $ E_0(M_L,M_R) = 2M_L \mu_L + 2M_R \mu_R$. The states $|l\rangle$ are high-energy states with two QPs, for example, $d_{k\uparrow}^+d_{q\downarrow}^+|M_L,M_R-1\rangle$ with the energy $E_k + \mu_L + E_q + \mu_R + 2M_L\mu_L + 2(M_R-1)\mu_R$. Thus, the second-order terms in the effective Hamiltonian correspond to virtual processes involving the creation of a pair of QPs followed by their annihilation. The resulting non-zero matrix elements are given by
\begin{equation}
\begin{split}
\langle M_L,M_R|H_\mathrm{eff}| M_L,M_R\rangle = E_0(M_L,M_R) + E_{0}^{(2)}  \\
\langle M_L-1,M_R+1|H_\mathrm{eff}|M_L,M_R\rangle = - T_\mathrm{CP}^{(2)} \\
\langle M_L+1,M_R-1)|H_\mathrm{eff}| M_L,M_R \rangle = - T_\mathrm{CP}^{(2)*}
\end{split}
\label{eq:matrixelements}
\end{equation}
where
\begin{equation}
E_0^{(2)} = - \sum_{kq\sigma}|T_{kq}|^2\left(\frac{|u_q|^2|v_k|^2}{E_k+E_q-eV} + \frac{|u_k|^2|v_q|^2}{E_k + E_q + eV}\right)
\end{equation}
is the energy shift, independent from $M_L$ and $M_R$, and where $eV = \mu_L -\mu_R$ denotes the bias voltage. In Eq. \eqref{eq:matrixelements}, we have also defined
\begin{equation}
T_\mathrm{CP}^{(2)} = \sum_{kq\sigma} T_{kq}^2 u_q v_k v_q^* u_k \frac{E_k + E_q}{(E_k+E_q)^2 - (eV)^2}
\label{eq:EJ}
\end{equation}
describing the amplitude of CP tunneling in the effective model. Note that the freedom in choosing the phase of $v_k$ and $v_q$ can be used to set $T_\mathrm{CP}^{(2)}$ to be real. In the following, we adopt this convention and use the standard notation $T_\mathrm{CP}^{(2)} = E_J /2$, where $E_J$ denotes the Josephson energy.

Based on the matrix elements, the effective Hamiltonian can be written as
\begin{equation}
H_\mathrm{eff} = H_0 + E_0^{(2)} - \frac{E_J}{2}\left(S_R^+ S_L + h.c.\right)
\end{equation}
which, using the phase representation of the operators $S_L=e^{i\hat{\varphi}_L}$ and $S_R=e^{i\hat{\varphi}_R}$, becomes
\begin{equation}
H_\mathrm{eff} = H_0 + E_0^{(2)} - E_J \cos\hat{\varphi}
\label{eq:Heff0qp2ord}
\end{equation}
where $\hat{\varphi}=\hat{\varphi}_L - \hat{\varphi}_R$ which forms a conjugate pair with the operator $\frac{1}{2}\sum_{M_L,M_R}(M_L-M_R)|M_L,M_R\rangle\langle M_L,M_R|$. 

Our result in Eq. \eqref{eq:Heff0qp2ord} together with Eq. \eqref{eq:EJ} essentially reproduces the findings of conventional perturbative approaches \cite{josephson1962, ambegoakar1963}. As for the voltage dependence of the Josephson energy observed in the denominator of Eq. \eqref{eq:EJ}: in typical JJ experiments, this voltage dependence represents a minor correction (see Appendix \ref{sec:constDOS}) and, hence, can be neglected in practical applications. The SWT approach has the advantage of explicitly showing that a unitary transformation exists between the Hamiltonian with single-electron tunneling and the conventional circuit-QED Hamiltonian with a $-E_J \cos\hat{\varphi}$ term. In principle, this unitary transformation should be applied to all operators analyzed within the framework of the Josephson Hamiltonian.
\jav{Another benefit of the SWT is that it directly provides operator expressions, unlike conventional perturbation expansions that results in corrections only to the energy and the wavefunction.}

An additional advantage of our approach is its systematic extensibility to more complex scenarios. The derivation presented above is based on two principal assumptions: the adequacy of a second-order expansion in the tunneling and the restriction of the low-energy subspace to QP-free states. These assumptions can be relaxed independently. In the following section, we expand the low-energy subspace by including the states that contain a single QP. In Sec. \ref{sec:4order}, we maintain the original low-energy QP-free subspace but increase the order of the perturbative expansion to fourth order.

\section{Extension to states with one quasiparticle}
\label{sec:1qp}

\subsection{Hilbert-space partitioning}

We modify the SWT presented in the previous section by extending the low-energy subspace with the states including one QP (1-QP states), such as $|k\sigma,M_L,M_R\rangle = d_{k\sigma}^+ |M_L,M_R \rangle$ with the energy $E_k +\mu_L+ E_0(M_L,M_R)$. These states are the lowest-energy states for an odd number of particles. It is also important to note that the effective Hamiltonian has no matrix elements between QP-free and 1-QP states; these states live in entirely separate Hilbert spaces.

Following Ref. \cite{bravyi2011}, we decompose the tunneling Hamiltonian into block-diagonal and block-off-diagonal terms, $V=V_{d} + V_{od}$. The block-off-diagonal part, $V_{od}$, includes the terms marked by $\fbox{$\rightarrow/\leftarrow A/C$}$ in Eq. \eqref{eq:Ht} which are responsible for transitions between the low- and high-energy subspaces. The block-diagonal part, $V_d$, consists of the remaining terms that describe transitions within the subspaces.

We note that the partitioning of the Hilbert space adopted in this section leads to "low-energy" and "high-energy" subspaces whose spectra actually overlap. In the specific case, a 1-QP state from the "low-energy" subspace with a very high energy, which is possible because the spectrum of the QPs is unbounded from above, can have the same energy as, for instance, a 3-QP state from the "high-energy" subspace composed from multiple low-energy excitations. The main condition for splitting the Hilbert space according to Ref. \cite{bravyi2011} is that there must exist a positive energy $\Delta E$ such that $|E_h - E_l|>\Delta E$ for all high-energy states $h$ and all low-energy states $l$. Our model does not fulfill this requirement. However, this condition is actually not a necessary but merely a sufficient condition for performing a proper SWT. The SWT remains applicable even when the condition $|E_h - E_l|>\Delta E$ is fulfilled only for states which are directly connected by the block-off-diagonal part of the perturbation. This requirement is, in fact,  fulfilled in our model.
Furthermore, this is in line with the physical intuition according to which the 1-QP states with very high  energies, i.e., far from the gap edge, are less relevant in describing the dominant behavior of the junction. All of these considerations justify the Hilbert space partitioning into sectors of different quasiparticle numbers.

\subsection{Effective Hamiltonian}

Using Eq. (3.13) of Ref. \cite{bravyi2011} and expanding to second order in the tunneling, we obtain the effective Hamiltonian
\begin{equation}
H_\mathrm{eff} = H_0 + V_{d} + H_{\mathrm{eff}}^{(2)}
\end{equation}
where the second order term $H_{\mathrm{eff}}^{(2)}$ has matrix elements
\begin{equation}
\begin{split}
\langle n|H_{\mathrm{eff}}^{(2)}|m\rangle &= \frac{1}{2}\sum_{l}\langle n|V_{od}|l\rangle\langle l |V_{od}|m\rangle \times \\
\times & \left(\frac{1}{E_n - E_l} + \frac{1}{E_m - E_l}\right).
\end{split}
\label{eq:Heff2toeval}
\end{equation}
Here, the relevant high energy states, $|l\rangle$, are 3-QP states, for example, $|k'\bar{\sigma}',k\sigma,q\sigma',M_L,M_R\rangle$ with the energy $E_k' + E_k + E_q + 2\mu_L + \mu_R + E_0(M_L,M_R)$.

We evaluate the sums in $H_{\mathrm{eff}}^{(2)}$ for all relevant processes (see Appendix \ref{sec:1qpprocesses} for details) and obtain the effective Hamiltonian
\begin{equation}
\begin{split}
H_{\mathrm{eff}} &= H_0 + E_0^{(2)} - E_J\cos\hat{\varphi}  + \\
& + \sum_{kq\sigma}\Big[\Big(T_{kq} u_k u_q - T_{kq}^* v_q v_k^* e^{-i\hat{\varphi}}\Big) d_{q\sigma}^+ d_{k\sigma} + h.c.\Big] + \\
&+ \sum_{kk'\sigma} t_{L,kk'}d_{k'\sigma}^+ d_{k\sigma} +
\sum_{kk'\sigma}\left(h_{L,kk'} e^{i\hat{\varphi}} d_{k'\sigma}^+ d_{k\sigma} + h.c.\right)  + \\ &+
\sum_{qq'\sigma} t_{R,qq'}d_{q'\sigma}^+ d_{q\sigma} + \sum_{qq'\sigma}\left(h_{R,qq'} e^{i\hat{\varphi}} d_{q'\sigma}^+ d_{q\sigma} + h.c.\right)
\label{eq:Heff1qp}
\end{split}
\end{equation}
which is one of the main results in this work. Compared to Eq. \eqref{eq:Heff0qp2ord}, we find several new features:
\begin{enumerate}
\item 
first-order terms, arising from $V_d$, corresponding to direct quasiparticle tunneling between the two leads;

\item 
new second-order terms that account for intralead QP scattering processes, either unassisted or assisted by CP tunneling.
\end{enumerate}
For instance, the term $h_{L,kk'} e^{i\hat{\varphi}} d_{k'\sigma}^+ d_{k\sigma}$ describes scattering of the QP from mode $k$ to mode $k'$, accompanied by a CP transfer from left to right. This correlated motion of QPs and CPs has implications for the eigenstates of the Hamiltonian and the resulting dynamics.

\jav{Furthermore, from the QP perspective, the intralead scattering terms of Eq. \eqref{eq:Heff1qp} can also be relevant for QPs thermalizing within the lead.} 

\subsection{Toy model}

\begin{figure}[!htbp]
\centering
\includegraphics[width=7cm]{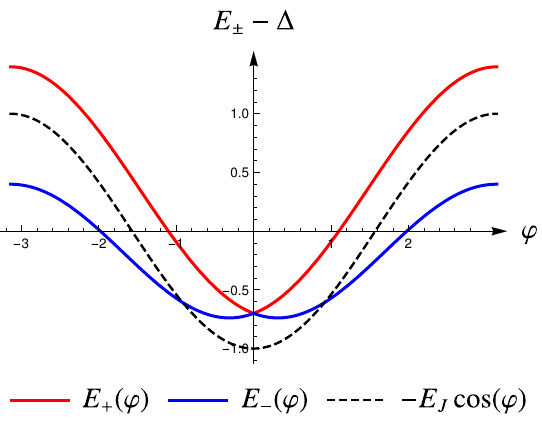}
\caption{Spectrum of the Josephson junction with a single quasiparticle. The parameters were set to $T=0.5$ and $h=t=0.1$ in units of $E_J$. The dashed curve shows the spectrum of the quasiparticle-free case. In reality, the quasiparticle-free and one-quasiparticle branches are separated by $\Delta$ which is not displayed in the plot since it greatly exceeds all other energy scales.}
\label{fig:JJspectrum}
\end{figure}

To further analyze the effects of the QP-CP coupling, we investigate a minimal toy model with a single energy level at the gap edge in each lead. For simplicity, we take identical gaps, such that the single QP energies are $E_k=E_q=\Delta$. We further assume $h=h_{Lkk}=h_{Rqq}$ and $t=t_{Lkk}=t_{Rqq}$. The amplitude of the direct tunneling, denoted by $T$, is assumed to be real.
In addition, we employ the phase eigenstate representation of the 1-QP states, defined as
\begin{equation}
\begin{split}
|k\sigma,M;\varphi\rangle = \sum_{m=-\infty}^{\infty}e^{i\varphi m}|k\sigma,M;m\rangle \\
|q\sigma,M;\varphi\rangle = \sum_{m=-\infty}^{\infty}e^{i\varphi m}|q\sigma,M;m\rangle
\end{split}
\end{equation}
where $|k\sigma,M;m\rangle$ is the same state as $|k\sigma,M_L,M_R\rangle$ with $M=M_L+M_R$ and $m=(M_R-M_L)/2$. On this two-dimensional Hilbert space, the effective Hamiltonian is represented by the matrix
\begin{gather}
\uu{H}_{\mathrm{eff}} = \Delta + t -\tilde{E}_J\cos\varphi + 
\left[ \begin{array}{cc} 0 & \frac{T}{2}\left(1-e^{-i\varphi}\right) \\ \frac{T}{2}\left(1 - e^{i\varphi}\right) & 0 \end{array} \right]
\label{eq:uuH1}
\end{gather}
where $\tilde{E}_J = E_J-2h$ is the renormalized Josephson energy. The eigenvalues of the matrix are calculated as
\begin{gather}
E_{\pm}(\varphi)= -\tilde{E}_J \cos{\varphi} + \Delta + t \pm T\left|\sin\left(\frac{\varphi}{2}\right)\right|
\end{gather}
which we plot in Fig. \ref{fig:JJspectrum} as a function of the phase difference. The existence of two solutions corresponds to the 2-dimensional fermionic Hilbert space of this toy model.

\jav{The energy functions $E_{\pm}(\varphi)$ have important practical consequences on the excitation frequency of a JJ-based qubit. }
In the absence of QPs, the energy landscape has the familiar form of $-E_J\cos\varphi$ with a minimum at $\varphi=0$. Near the minimum, the function is approximated with the parabola $-E_J + \varphi^2 E_J/4$ where the curvature $E_J/4$ directly determines the qubit excitation frequency. The presence of the QP introduces a vertical shift of the spectrum by $\Delta$. Since $\Delta$ is typically much larger than the other energy scales, it ensures a wide separation between the QP-free and the 1-QP spectra. The parameter $t$ produces only a minor vertical shift without modifying the overall shape of the curve. The tunneling amplitude $T$ shifts the spectrum horizontally while the parameter $h$ renormalizes $E_J$ as $\tilde{E}_J = E_J -2h$. The resulting curve has its minimum displaced to $\varphi_0 = \pm \arcsin[T/(2\tilde{E}_J)]$ and the parabola around the minimum takes the form
\begin{equation}
E_-(\varphi)\approx -\tilde{E}_J-\frac{T^2}{4\tilde{E}_J} + \frac{\tilde{E}_J}{4}\left(1-\frac{T^2}{4\tilde{E}_J^2}\right) (\varphi-\varphi_0)^2
\end{equation}
\jav{with the modified curvature $\frac{\tilde{E}_J}{4}\left(1-\frac{T^2}{4\tilde{E}_J^2}\right)$.}
The change in curvature is a clear indication that the resulting dynamics differ from the QP-free case \jav{and also implies that the presence of QPs around the JJ modifies the qubit excitation frequency.} \jav{To explore how exactly the excitation frequency changes due to this effect, a full circuit QED model of the junction would be required by} including the charging energy \jav{term in the total Hamiltonian}. This, however, lies beyond the scope of the present work.

\jav{The toy model is based on the assumption that, for quasiparticles, only a single energy level at the gap edge is considered. In a realistic situation, multiple quasiparticle levels may be relevant leading to different energy landscapes but the same qualitative features, namely the modification of qubit excitation frequency and the offset of the minimal energy in the $\varphi$ space, are expected to appear.}

\section{Fourth-order SWT}
\label{sec:4order}
The SWT can also be extended to higher order in the tunneling amplitude in a systematic manner. In the present section, we consider again the low-energy subspace consisting of the QP-free states only. 

To carry out the SWT to fourth order, we assume that the transformation operator $R$ contains terms of first and third order in the tunnelling, $R=R^{(1)} + R^{(3)}$. We extend Eq.~\eqref{eq:expansion2} by expanding the effective Hamiltonian to fourth order in the tunnelling. Further details are provided in Appendix~\ref{sec:fourthprocesses}. The first-order term $R^{(1)}$ is chosen so that the linear terms cancel in the effective Hamiltonian, leading to the condition $V + [R^{(1)},H_0]=0$. Similarly, the third-order contribution $R^{(3)}$ is determined such that the third-order terms of the effective Hamiltonian also vanish, yielding the condition
\begin{equation}
- \frac{1}{3}\left[R^{(1)},\left[R^{(1)},\left[R^{(1)},H_0\right]\right]\right] + \left[R^{(3)},H_0\right] = 0.
\end{equation}
With these conditions, the fourth-order effective Hamiltonian takes the form
\begin{equation}
H_\mathrm{eff}=H_0 + E_0^{(2)} - E_J\cos\hat{\varphi} + H_{\mathrm{eff}}^{(4)}.
\label{eq:Heff4ord}
\end{equation}
%
Analysis of the fourth order processes, \jav{see the Appendix \ref{sec:4order}}, shows that $H_\mathrm{eff}^{(4)}$ acts on the quasiparticle-free Hilbert space as 
\begin{equation}
H_\mathrm{eff}^{(4)} = E_0^{(4)} + T^{(4)}_\mathrm{CP}\cos\hat{\varphi} + D^{(4)}\cos(2\hat{\varphi}).
\end{equation}
Here, the constant term $E_0^{(4)}$ and the single CP tunneling term $T^{(4)}_\mathrm{CP}\cos\hat{\varphi}$ can be absorbed into the second-order terms, leading to a redefinition of $E_0^{(2)}$ and $E_J$. The new non-trivial term  $D^{(4)}\cos(2\hat{\varphi})$, referred to as the second Josephson harmonic in the literature~\cite{secondharmonic2007,Willsch2024}, describes the simultaneous tunneling of two CPs. \jav{This term originates from fourth-order processes involving the creation and annihilation of two QPs, accompanied in some order by the creation of two CPs in one lead and the annihilation of two CPs in the other. For example, the process $\fbox{$\rightarrow C$}\fbox{$\rightarrow C$}\fbox{$\rightarrow A$}\fbox{$\rightarrow A$}$ effectively describes the tunneling of two CPs from the left lead to the right one. } By taking into account even higher orders in the SWT, higher harmonics can similarly be captured.

The parameters $T^{(4)}_\mathrm{CP}$ and $D^{(4)}$ can be calculated \jav{for a specific microscopic model of the leads and the weak link} as detailed in Appendix~\ref{sec:fourthprocesses}.
\jav{For leads with constant normal-state density of states, $\mathcal{D}$, and a junction with momentum-independent tunneling amplitude $T$}, the coefficients \jav{ $T^{(4)}_\mathrm{CP}$ and $D^{(4)}$ are given in Appendix \ref{sec:constDOS}. Note that the momentum-independence corresponds to a tunneling between one distinguished level in the left lead, $\sim\sum c_{k\sigma}$, and one in the right lead, $\sim \sum c_{q\sigma}$, effectively describing a one-channel tunneling (rank-one coupling matrix). Neglecting constant terms $E_0$, the eigenvalues of the effective Hamiltonian \eqref{eq:Heff4ord} are obtained as
\begin{equation}
\begin{split}
E(\varphi) = \left(-E_J + \frac{2T^4\mathcal{D}^2}{\Delta}\left(\frac{\pi^2}{16} -\alpha\mathcal{D}^2\Delta^2     \right)\right)\cos\varphi + \\ + \frac{2T^4\mathcal{D}^2}{\Delta}\left( \alpha' -\alpha''\mathcal{D}^2 \Delta^2    \right)\cos(2\varphi),
\end{split}
\label{eq:Efi}
\end{equation}
where $\alpha'\approx 0.0896$, $\alpha''\approx 2.307$ and $\alpha$ is a positive numerical constant depending on the high-energy cutoff $W$. The numerical constants depend on $\mathcal{D}\Delta$, which is a small quantity in the conventional case, where $\alpha$ and $\alpha''$ terms represent a tiny correction, however in flat-band superconductors they could significantly affect the overall parameter values. Eq.~\eqref{eq:Efi} shows that in the conventional case by increasing the tunneling amplitude $T$, the first harmonic is suppressed while the second harmonic is enhanced. This is in accordance with the results of the mesoscopic study in Ref. \cite{Willsch2024} as the channel transparency is related to the tunneling amplitude. For very flat bands, the trend could become reversed.

Furthermore, we find that, as $\mathcal{D}\Delta$  increases, the fourth-order contribution to the first harmonic, $T_{CP}^{(4)}$, becomes less significant compared to $D^{(4)}$, see the inset in Fig.~\ref{fig:harmonics}.

\begin{figure}[h]
\centering
\includegraphics[width=8cm]{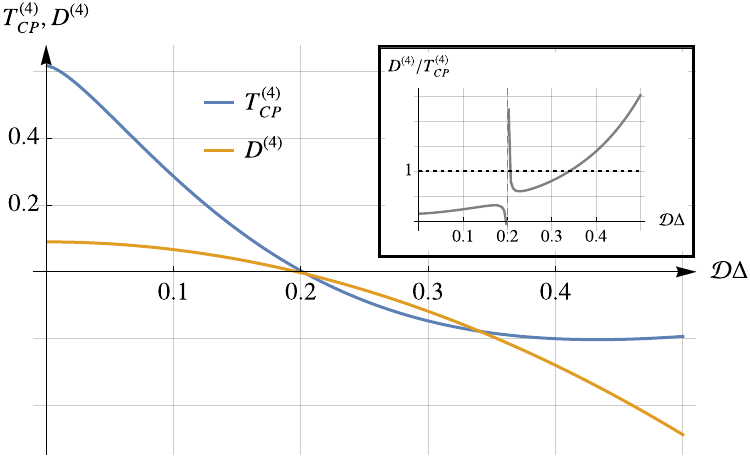}
\caption{The fourth-order contributions to the first and second Josephson harmonics as a function of $\mathcal{D}\Delta$. The parameters $T_{CP}^{(4)}$ and $D^{(4)}$ are measured in units of $T^4\mathcal{D}^2/\Delta$. For the plot, constant density of states and momentum-independent tunneling amplitude is assumed. Furthermore, the bandwidth is assumed to be $W = \sqrt{(1/\mathcal{D})^2 + \Delta^2}$, where $\mathcal{D}$ is the normal-state density of states. Inset: the ratio of the two contributions indicating that larger $\mathcal{D}\Delta$ generates more significant second harmonic.}
\label{fig:harmonics}
\end{figure}

Note that Eq. \eqref{eq:Efi} is valid only in the perturbative regime, $T\ll \Delta$. The second harmonic cannot dominate over the first harmonic within this framework. To design a junction dominated by second harmonic behavior, the tunneling must be constructed such that the amplitude of the first harmonic vanishes. This can be achieved by considering, for example, a two-channel tunneling with $T_{kq} = T_1 \gamma_{1q}^*\gamma_{1k} + T_2 \gamma_{2q}^*\gamma_{2k}$ (rank-two matrix) instead of the momentum-independent tunneling amplitudes. Here, the tunneling is realized in two channels, one between $\sum \gamma_{1k}c_{k\sigma}$ and $\sum \gamma_{1q}c_{q\sigma}$ and another one between $\sum \gamma_{2k}c_{k\sigma}$ and $\sum \gamma_{2q}c_{q\sigma}$. 

For the tunneling amplitudes of the two channels, we define $T_1=T$ and $T_2=rT$ and we assume that $T$ is real. The second-order term in the effective Hamiltonian, $-E_J\cos\varphi$, is a quadratic function of $r$ and cancels due to destructive interference if the ratio $r$ of the tunneling amplitudes is chosen as 
\begin{equation}
r=\frac{T_2}{T_1} = \frac{-A_{12} \pm\sqrt{A_{12}^2-A_{11}A_{22}}}{A_{22}}:=r_0
\label{eq:T12ratio}
\end{equation}
where both $+$ and $-$ may produce a valid solution, and
\begin{equation}
A_{jj'}=\sum_{kq}\gamma_{jq}^{*}\gamma_{jk}\gamma_{j'q}^{*}\gamma_{j'k}\frac{\Delta^3}{E_kE_q(E_k+E_q)}
\end{equation}
depends on microscopic details. By setting the ratio of Eq. \eqref{eq:T12ratio}, only fourth-order terms remain non-zero in the effective Hamiltonian. 

The ratio $r$ can be further fine-tuned such that $E_J - 2T_{CP}^{(4)}$ vanishes, i.e., the first harmonic is completely canceled. Since $T_{CP}^{(4)}$ is a quartic polynomial in $r$, analytical solution of the equation $E_J - 2T_{CP}^{(4)}=0$ is difficult to obtain. Nevertheless, an approximate solution can be derived up to second order in $T^2/\Delta^2$. By introducing the function $f(r)$ via $T_{CP}^{(4)}=\frac{T^4}{2\Delta^3}f(r)$, we obtain that the ratio of the two channel tunneling amplitudes should be chosen as
\begin{equation}
r = r_0 + \frac{T^2}{2\Delta^2}\frac{f(r_0)}{A_{22} r_0 + A_{12}}
\end{equation} to ensure that $E_J - 2T_{CP}^{(4)}$ vanishes up to fourth order in $T/\Delta$, leaving a purely second harmonic junction to this order.
}

\section{Discussion}
The effective Hamiltonian of a JJ, consisting of two superconducting leads separated by a thin insulating layer, has been derived in various contexts of current interest. The conventional Hamiltonian $-E_J\cos\hat{\varphi}$ was obtained from the charge-conserving BCS theory by a second-order SWT under the condition that the low-energy subspace consists solely of quasiparticle-free states. Although the final result of this derivation is not novel, the method of SWT represents a new and systematic approach to obtaining effective JJ Hamiltonians and provides new insights into the JJ phenomena. First, it highlights that the effective Hamiltonian $-E_J\cos\hat{\varphi}$ is linked by a unitary transformation to the single-electron tunneling model. This unitary transformation may also be relevant to other physical quantities and the transport properties of the junction. Second, this procedure avoids the complications of requantizing the phase of the order parameter, which typically arise in derivations based on the Ginzburg-Landau theory, since the phase difference here naturally emerges as an operator. 

Taking the advantage of systematic extensibility to more complex situations,
in this work we derived the effective Hamiltonian for the scenario in which a single quasiparticle is present in the ground state of the system. We found that this modification introduces intralead scattering terms into the effective Hamiltonian. Some of these terms are assisted by Cooper-pair tunneling, thereby describing correlated motion of the quasiparticles and Cooper pairs. By analyzing a toy model with a single one-particle state in each lead, we showed that this effect modifies the conventional $-E_J\cos\hat{\varphi}$ energy landscape and, therefore, can alter the excitation energy of a superconducting qubit poisoned by a quasiparticle. We believe that this finding represents an important step toward a comprehensive effective model of Josephson-junction-based qubits, which may be completed by including the charging energy in the description.

The SWT also provides an opportunity to analyze higher-order corrections and Josephson harmonics. We analyzed the fourth-order terms on the low-energy subspace of quasiparticle-free states and obtained second Josephson harmonics characterized by the $\cos(2\hat{\varphi})$ term. \jav{We demonstrated how this approach can be applied to engineer the ratio between the amplitudes of the first and second harmonic by showcasing two simplest situations, rank-one and rank-two tunneling (overlap) matrices. In particular, for two scattering channels we have established that the fundamental $\cos\phi$ can be suppressed through interference.}

\begin{acknowledgments}
This work was supported by the National Research, Development and Innovation Office - NKFIH  Project Nos. K134437 and K142179. We acknowledge the support of the Slovenian Research and Innovation Agency (ARIS) under P1-0416 and J1-3008.
\end{acknowledgments}

\section*{Data availability statement}
The data supporting the findings are available in the Supplementary Material \cite{supplscnfl}.

\bibliographystyle{apsrev}
\bibliography{jjdot}

\appendix
\section{Second order processes in the presence of one quasiparticle state}
\label{sec:1qpprocesses}
We analyze the second-order processes starting from a state with one QP in the left lead, $|k\sigma,M_L,M_R\rangle$, and evaluate the last term of Eq. \eqref{eq:Heff2toeval} in the main text.
The relevant processes are listed in Tab. \ref{tab:1qpprocessestrue}. The notation /R refers to the process when the exact same QP is annihilated that was created before. The notation /S represents a process when a different QP is annihilated and so describing effective scattering of the QP. We note that $H_\mathrm{eff}^{(2)}$ has matrix elements only between 1-QP states with a QP located in the same lead. Namely, only $\langle k|H_\mathrm{eff}^{(2)}|k'\rangle$ are non-zero, while all $\langle k|H_\mathrm{eff}^{(2)}|q\rangle$ matrix elements vanish. Furthermore, no spin-flip processes are present.

The relevant high energy states, $|l\rangle$, are the 3-QP states such as $|k'\bar{\sigma}',k\sigma,q\sigma',M_L,M_R\rangle$ with the energy $E_k' + E_k + E_q + 2\mu_L + \mu_R + E_0(M_L,M_R)$.

\begin{widetext}
\begin{center}
\begin{table}[h]
\small
\setlength{\arrayrulewidth}{0.5pt}
\begin{tabular}{|c|c|c|c|}
\hline
Process & $|l\rangle$ & $|n\rangle$ & $\langle n|H_{T,od}|l\rangle\cdot\langle l|H_{T,od}|m\rangle$ \\
\hline
\multicolumn{4}{|c|}{Processes starting from the left lead: $|m\rangle = |k\sigma,M_L,M_R\rangle$} \\
\hline
$\fbox{$\rightarrow C$}$ $\fbox{$\leftarrow A$}$/R & $|k'\bar{\sigma}',k\sigma,q\sigma',M_L-1,M_R\rangle$ & $|k\sigma,M_L,M_R\rangle$ & $|T_{k'q}|^2u_q^2 |v_{k'}|^2$ \\
$\fbox{$\rightarrow C$}$ $\fbox{$\leftarrow A$}$/S & $|k'\bar{\sigma}',k\sigma,q\sigma',M_L-1,M_R\rangle$ & $|k'\sigma,M_L,M_R\rangle$ & $-\delta_{\bar{\sigma}'\sigma}T_{k'q}T^*_{kq}  u_q^2 v_{k'}v_k^*$ \\
$\fbox{$\rightarrow C$}$ $\fbox{$\rightarrow A$}$/R & $|k'\bar{\sigma}',k\sigma,q\sigma',M_L-1,M_R\rangle$ & $|k\sigma,M_L-1,M_R+1\rangle$ & $T_{k'q}^2u_q v_q^* u_{k'} v_{k'}$ \\
$\fbox{$\rightarrow C$}$ $\fbox{$\rightarrow A$}$/S & $|k'\bar{\sigma}',k\sigma,q\sigma',M_L-1,M_R\rangle$ & $|k'\sigma,M_L-1,M_R+1\rangle$ & $-\delta_{\bar{\sigma}'\sigma}T_{k'q}T_{kq}  u_q v_q^* u_{k}v_{k'}$ \\
$\fbox{$\leftarrow C$}$ $\fbox{$\rightarrow A$}$/R & $|k'\bar{\sigma}',k\sigma,q\sigma',M_L,M_R-1\rangle$ & $|k\sigma,M_L,M_R\rangle$ & $|T_{k'q}|^2 u_{k'}^2 |v_{q}|^2$ \\
$\fbox{$\leftarrow C$}$ $\fbox{$\rightarrow A$}$/S & $|k'\bar{\sigma}',k\sigma,q\sigma',M_L,M_R-1\rangle$ & $|k'\sigma,M_L,M_R\rangle$ & $-\delta_{\bar{\sigma}'\sigma}T_{k'q}^* T_{kq} u_k u_{k'} |v_{q}|^2$ \\
$\fbox{$\leftarrow C$}$ $\fbox{$\leftarrow A$}$/R & $|k'\bar{\sigma}',k\sigma,q\sigma',M_L,M_R-1\rangle$ & $|k\sigma,M_L+1,M_R-1\rangle$ & $(T_{k'q}^*)^2 u_{k'}v_{k'}^* u_q v_{q}$ \\
$\fbox{$\leftarrow C$}$ $\fbox{$\leftarrow A$}$/S & $|k'\bar{\sigma}',k\sigma,q\sigma',M_L,M_R-1\rangle$ & $|k'\sigma,M_L+1,M_R-1\rangle$ & $-\delta_{\bar{\sigma}'\sigma}T_{k'q}^*T_{kq}^* u_{k'}v_{k}^* u_q v_{q}$ \\
\hline
\end{tabular}
\caption{Matrix elements in the 1-qp sectors}
\label{tab:1qpprocessestrue}
\end{table}
\end{center}

The relevant matrix elements of $H_\mathrm{eff}^{(2)}$ are obtained as follows.

\begin{equation}
\langle k\sigma, M_L, M_R|H_\mathrm{eff}^{(2)}|k\sigma,M_L,M_R\rangle = E_0^{(2)} + \underbrace{\sum_{q}|T_{kq}|^2\left(\frac{u_q^2 |v_k|^2}{E_q + E_k - \mu_L + \mu_R} +
\frac{u_k^2 |v_q|^2}{E_q + E_k - \mu_R + \mu_L}\right)}_{t_{L,kk}}
\end{equation}

\begin{equation}
\langle k\sigma,M_L-1, M_R + 1|H_\mathrm{eff}^{(2)}|k\sigma,M_L,M_R\rangle =  - \frac{E_J}{2} + \underbrace{\sum_q T_{kq}^2 u_q v_q^* u_k v_k\frac{E_k + E_q}{(E_k + E_q)^2 - (\mu_L - \mu_R)^2}}_{h_{L,kk}}
\end{equation}

\begin{equation}
\langle k\sigma,M_L+1,M_R-1|H_\mathrm{eff}^{(2)}|k\sigma,M_L,M_R\rangle = - \frac{E_J}{2} + h_{L,kk}^*
\end{equation}

Furthermore, if $k'\neq k$, we have
\begin{equation}
\begin{split}
t_{L,kk'} &:= \langle k'\sigma, M_L, M_R|H_\mathrm{eff}^{(2)}|k\sigma,M_L,M_R\rangle = \\ & =\frac{1}{2}\sum_q \left[T_{kq}T_{k'q}^* u_k u_{k'}
|v_q|^2 \left(\frac{1}{E_q + E_k + \mu_L-\mu_R} + \frac{1}{E_q + E_{k'} +\mu_L-\mu_R}\right) + \right. \\
& \left. + T_{kq}^* T_{k'q}v_k^* v_{k'}
u_q^2 \left(\frac{1}{E_q + E_k - \mu_L+ \mu_R} + \frac{1}{E_q + E_{k'} - \mu_L+ \mu_R}\right) \right]
\end{split}
\label{eq:tLkkp}
\end{equation}

\begin{equation}
\begin{split}
h_{L,kk'} & := \langle k'\sigma, M_L-1, M_R+1|H_\mathrm{eff}^{(2)}|k\sigma,M_L,M_R\rangle  = \\ & = \frac{1}{2}\sum_q T_{kq}T_{k'q}u_q v_q^* u_k v_{k'}\left( \frac{1}{E_q + E_{k'} - (\mu_L - \mu_R)} + \frac{1}{E_q + E_k + \mu_L - \mu_R}\right)
\end{split}
\label{eq:hL}
\end{equation}

\begin{equation}
\langle k'\sigma,M_L+1,M_R-1|H_\mathrm{eff}^{(2)}|k\sigma,M_L,M_R\rangle = h_{L,k'k}^*
\end{equation}

The processes starting from a state with a single quasiparticle in the right lead, like $|q\sigma,M_L,M_R\rangle$, can similarly be described and evaluated. Therefore, the total effective Hamiltonian can be written as
\begin{equation}
\begin{split}
H_\mathrm{eff}^{(2)} & = E_0^{(2)} - \frac{E_J}{2}\left( S_R^+ S_L + S_L^+ S_R \right)  + \sum_{kk'\sigma} t_{L,kk'}d_{k'\sigma}^+ d_{k\sigma} +
\sum_{kk'\sigma}\left(h_{L,kk'} S_R^+ S_L d_{k'\sigma}^+ d_{k\sigma} + h.c.\right)  +  \\ &+
\sum_{qq'\sigma} t_{R,qq'}d_{q'\sigma}^+ d_{q\sigma} + \sum_{qq'\sigma}\left(h_{R,qq'} S_R^+ S_L d_{q'\sigma}^+ d_{q\sigma} + h.c.\right)
\end{split}
\end{equation}
which is identical to Eq. \eqref{eq:Heff1qp} in the main text. Note that this result acts on states with only one QP. In case of multipe QPs, the effective model is expected to include effective interactions between the QPs.
\end{widetext}
\vspace{1mm}

\section{Fourth-order SWT}
\label{sec:fourthprocesses}
We summarize the results of the fourth-order SWT with the transformation operator of $R=R^{(1)} + R^{(3)}$. Here, $R^{(1)}$ is linear in the perturbation while $R^{(3)}$ is proportional to $V^3$.
The fourth-order expansion of the effective Hamiltonian is obtained as
\begin{widetext}
\begin{equation}
\begin{split}
H_\mathrm{eff}&=e^R H e^{-R}\approx \underbrace{H_0}_{0^{\mathrm{th}}} + \underbrace{V + \left[R^{(1)},H_0\right]}_{1^{\mathrm{st}}} + \underbrace{\left[R^{(1)},V\right] + 
\frac{1}{2}\left[R^{(1)},\left[R^{(1)},H_0\right]\right]}_{2^{\mathrm{nd}}} +  \\ 
& + \underbrace{\frac{1}{2}\left[R^{(1)},\left[R^{(1)},V\right]\right] + 
\frac{1}{3!}\left[R^{(1)},\left[R^{(1)},\left[R^{(1)},H_0\right]\right]\right] + \left[R^{(3)},H_0\right]}_{3^{\mathrm{rd}}} + \\ 
& + \underbrace{\frac{1}{3!}\left[R^{(1)},\left[R^{(1)},\left[R^{(1)},V\right]\right]\right] + \frac{1}{4!}\left[R^{(1)},\left[R^{(1)},\left[R^{(1)},\left[R^{(1)},H_0\right]\right]\right]\right]}_{4^{\mathrm{th}}} + \\
& + \underbrace{\left[R^{(3)},V\right] + \frac{1}{2}\left[R^{(1)},\left[R^{(3)},H_0\right]\right] + \frac{1}{2}\left[R^{(3)},\left[R^{(1)},H_0\right]\right]}_{4^{\mathrm{th}}}.
\end{split}
\label{eq:SW5}
\end{equation}
\end{widetext}
The first-order terms cancel if $V + \left[R^{(1)},H_0\right] = 0$. Furthermore, we set $R^{(3)}$ to cancel the third-order terms leading to the condition
\begin{equation}
- \frac{1}{3}\left[R^{(1)},\left[R^{(1)},\left[R^{(1)},H_0\right]\right]\right] + \left[R^{(3)},H_0\right] = 0.
\label{eq:3def}
\end{equation}

Under these conditions, the effective Hamiltonian simplifies to
\begin{equation}
\begin{split}
H_\mathrm{eff}&= H_0 - \frac{1}{2} \left[R^{(1)},\left[R^{(1)},H_0\right]\right] + \\
&+\frac{1}{8}\left[R^{(1)},\left[R^{(3)},H_0\right]\right] - \frac{1}{2}\left[R^{(3)},\left[R^{(1)},H_0\right]\right]
\end{split}
\end{equation}
in accordance with Ref.~\cite{bravyi2011}.

The matrix elements of the operators $R^{(1)}$ and $R^{(3)}$ in the eigenbasis of $H_0|n\rangle = E_n|n\rangle$ are calculated as
\begin{widetext}
\begin{equation}
\begin{split}
&\langle n|R^{(1)}|m\rangle  = \frac{\langle n|V|m\rangle}{E_n - E_m}, \\
&\langle n|R^{(3)}|m\rangle  = \frac{1}{3(E_n-E_m)}  \sum_{kl}\frac{\langle n|V|k\rangle \langle k|V|l\rangle \langle l|V|m\rangle}{(E_n-E_k)(E_k-E_l)(E_l-E_m)}\left(E_n - 3E_k +3E_l - E_m\right).
\end{split}
\end{equation}
The matrix elements of the fourth-order contribution of the effective Hamiltonian is given by
\begin{equation}
\begin{split}
\langle n|H_{\mathrm{eff}}^{(4)} |m\rangle &=\frac{1}{8}\sum_{l_3 l_2 l_1}\frac{\langle n|V|l_3\rangle \langle l_3|V|l_2\rangle \langle l_2|V|l_1\rangle\langle l_1|V|m\rangle}{(E_n-E_{l_3})(E_{l_3}-E_{l_2})(E_{l_2}-E_{l_1})(E_{l_1}-E_m)}\times  \\ 
&\times\left(-E_n + 4 E_{l_3} - 6 E_{l_2} + 4 E_{l_1} - E_m + 4(E_n - E_m)\left( \frac{E_{l_2}-E_{l_1}}{E_{l_3}-E_m}-\frac{E_{l_3}-E_{l_2}}{E_{n}-E_{l_1}}\right)\right).
\end{split}
\end{equation}
\end{widetext}

Restricting the states $|n\rangle$ and $|m\rangle$ to the QP-free states, the fourth-order terms of the effective Hamiltonian have the following structure:
\begin{equation}
\begin{split}
H^{(4)}_{\mathrm{eff}} = E_{0}^{(4)} \jav{+} T^{(4)}_\mathrm{CP} S_R^+ S_L \jav{+} T^{(4)*}_\mathrm{CP} S_{L}^+ S_R + \\ 
+ D^{(4)} S_R^+ S_R^+ S_L S_L + D^{(4)*} S_L^+ S_L^+ S_R S_R . 
\end{split}
\end{equation}
The $T^{(4)}_\mathrm{CP}$ terms can be subsumed into the leading-order Josephson energy. The new term proportional to $\cos(2\hat{\varphi})$ describes simultaneous tunneling of two CP. The values of $T^{(4)}_\mathrm{CP}$ and $D^{(4)}$ can be computed from the microscopic properties of the leads. At zero bias voltage, we obtain the following exact expressions for these quantities:
\begin{widetext}
\begin{equation}
\begin{split}
T^{(4)}_\mathrm{CP} & = 8\sum_{kq} T_{kq}^2 \left| T_{kq}\right| ^2 u_{k} u_{q} v_{k}  v_{q}^*\frac{u_{k}^2 \left| v_{q}\right| ^2+u_{q}^2 \left| v_{k}\right| ^2}{(E_{k}+E_{q})^3 }
-   \\ 
& - 2\sum_{\substack{kq \\ k'q'}}T_{k'q'}T_{kq'}  u_{q'} v_{q'}^* \left(T_{k'q} T_{kq}^* u_{k}^2 u_{k'} v_{k'} \left| v_{q}\right| ^2  +T_{kq} T_{k'q}^* u_{q}^2 u_{k}v_{k} \left| v_{k'}\right| ^2 \right)\frac{E_k + E_q + E_{k'}+ E_{q'} }{(E_{k}+E_{q})(E_{k'}+E_{q'}) (E_{k}+E_{q'}) (E_{k'}+E_{q})},
\end{split}
\label{eq:T4}
\end{equation}

\begin{equation}
D^{(4)} = 4\sum_{kq}\frac{T_{kq}^4 u_{k}^2  u_{q}^2 v_{k}^2 \left(v_{q}^*\right)^2}{(E_{k}+E_{q})^3 } - 
 2\sum_{\substack{kq \\ k'q'}}\frac{T_{k'q} T_{k'q'} T_{kq} T_{kq'} u_{k} u_{k'} u_{q} u_{q'} v_{k} v_{k'} v_{q}^* v_{q'}^*}{(E_{k}+E_{q}) (E_{k}+E_{q'}) (E_{k'}+E_{q})}.
\label{eq:D4}
\end{equation}
\end{widetext}

\section{Parameters of the effective model in the case of constant density of states}
\label{sec:constDOS}
In this section, we evaluate the effective model parameters by assuming a constant normal-phase density of states in the superconducting leads, $\mathcal{D}_L(\xi_k) = \mathcal{D}_R(\xi_q) = \mathcal{D}$. Furthermore, we assume that both leads have the same real-valued gap $\Delta$, identical QP spectra and that the tunneling amplitude $T_{kq} = T$ is real and momentum-independent. Under these assumptions, the Josephson energy, as defined in Eq.~\eqref{eq:EJ}, is obtained as
\begin{equation}
\begin{split}
& E_J = 4 T^2\mathcal{D}^2 \Delta^2 \times \\
& \times \int_\Delta^\infty \frac{\mathrm{d}E}{\sqrt{E^2-\Delta^2}}   \int_\Delta^\infty \frac{\mathrm{d}E'}{\sqrt{E'^2-\Delta^2}} \frac{E+E'}{\left((E+E')^2 - \left(eV\right)^2\right)}
\end{split}
\end{equation}
which, in general, can be evaluated numerically only. 
At low bias voltages, $|eV|\ll\Delta$, we have
\begin{equation}
E_J = \pi^2T^2\mathcal{D}^2\Delta \left(1 + \frac{1}{16}\left(\frac{eV}{\Delta}\right)^2\right).
\end{equation}
We note that in a typical experimental setup, the correction from the bias voltage is expected to be very small and, hence, can be neglected.

The energy shift is obtained as
\begin{equation}
\begin{split}
& E_0^{(2)} = -4T^2\mathcal{D}^2 \Delta^2 \times \\
& \times \int_\Delta^\infty \frac{\mathrm{d}E}{\sqrt{E^2-\Delta^2}}   \int_\Delta^\infty \frac{\mathrm{d}E'}{\sqrt{E'^2-\Delta^2}} 
\frac{EE'(E+E')}{\left((E+E')^2 - \left(eV\right)^2\right)}
\end{split}
\end{equation}
which must be regularized by an ultra-violet cutoff, $W$, resulting in the leading term $E_0^{(2)}=-4\mathcal{D}^2T^2 W\ln\left(\frac{W}{\Delta}\right)$.

We also evaluate the parameter $t_{L,kk'}$ as defined in Eq. \eqref{eq:tLkkp},
\begin{equation}
\begin{split}
t_{L,kk'} & = \frac{T^2\mathcal{D}}{2}\left[u_k u_{k'} \left( f_t(E_k+eV) + f_t(E_{k'} + eV)\right) + \right. \\ & + \left. v_k^* v_{k'} \left( f_t(E_k-eV) + f_t(E_{k'} - eV)\right)\right]
\end{split}
\end{equation}
where 
\begin{equation}
f_t(E) = \int_\Delta^\infty\frac{\mathrm{d} E'\, E'}{\sqrt{E'^2-\Delta^2}}\frac{1}{E'+E}
\end{equation}
which is regularized by an ultraviolet cut-off, $W$, and to leading order in $\Delta/W$ and close to the gap edge, we obtain 
\begin{equation}
f_t(E)\approx -1-\frac{2}{3}\frac{E-\Delta}{\Delta} + \ln\left(\frac{2W}{\Delta}\right).
\end{equation}

Furthermore, we calculate $h_{L,kk'}$ \eqref{eq:hL},
\begin{equation}
h_{L,kk'}=\frac{T^2 \mathcal{D} u_kv_{k'}}{2}\left( f_h(E_{k'} - eV) + f_h(E_k + eV)\right)
\end{equation}
where
\begin{equation}
f_h(E) = \int_\Delta^\infty \frac{\mathrm{d} E'\, \Delta}{\sqrt{E'^2-\Delta^2}}\frac{1}{E'+E}=\frac{\Delta\cdot\mathrm{arcosh}\left(\frac{E}{\Delta}\right)}{\sqrt{E^2-\Delta^2}}
\end{equation}
which, close to the gap edge, is approximated by $f_h(E)\approx 1- (E-\Delta)/3\Delta$.

We also evaluate the parameters of the fourth-order processes. The coefficient $T_{Cp}^{(4)}$, as defined in Eq. \eqref{eq:T4}, is calculated as
\begin{equation}
T^{(4)}_\mathrm{CP} = \jav{\frac{\pi^2}{16}}\frac{T^4\mathcal{D}^2}{\Delta} - \alpha T^4\mathcal{D}^4 \Delta
\end{equation}
where
\begin{equation}
\begin{split}
\alpha = & \int_1^\infty \int_1^\infty \int_1^\infty \int_1^\infty \frac{\mathrm{d}y \mathrm{d}y' \mathrm{d}x \mathrm{d}x'}{\sqrt{y^2-1}\sqrt{y'^2-1}\sqrt{x^2-1}\sqrt{x'^2-1}} \times\\ 
& \times \frac{(x+x')(y+y')(x+x'+y+y')}{(x+y')(x' + y)(x + y)(x' + y')}
\end{split}
\end{equation}
is a numerical constant. Note that the integral diverges, so the integrals shall be regularized by setting the upper limits to $W/\Delta$. \jav{By setting $W/\Delta = 1000$, the numerical integral results in $\alpha\approx 356.6$.}

The parameter $D^{(4)}$, as defined in Eq. \eqref{eq:D4}, is computed as
\begin{equation}
D^{(4)}=\alpha'\frac{T^4\mathcal{D}^2}{\Delta} - \alpha'' T^4\mathcal{D}^4 \Delta 
\end{equation}
where
\begin{equation}
\alpha' = \int_1^\infty\int_1^\infty\frac{\mathrm{d}x\mathrm{d}y}{\sqrt{x^2-1}\sqrt{y^2-1}}\frac{1}{xy(x+y)^3}\approx 0.0896
\end{equation}
\begin{equation}
\alpha'' = \jav{2}\int_1^\infty\int_1^\infty\frac{\mathrm{d}x\mathrm{d}y}{(x^2-1)(y^2-1)}\frac{\mathrm{arcosh}(x)\mathrm{arcosh}(\jav{y})}{(x+y)}\approx 2.307
\end{equation}
are numerical constants.

\jav{
The energy spectrum of the fourth order effective Hamiltonian is computed as
\begin{equation}
E(\varphi) = E_0^{(2)} + E_0^{(4)} + \left(-E_J + 2T_{CP}^{(4)}\right)\cos\varphi + 2D^{(4)}\cos(2\varphi)
\end{equation}
where $\varphi$ is the eigenvalue of the operator $\hat{\varphi}$ defined through $S_R^+S_L =e^{i\hat{\varphi}}$. By omitting the constant terms $E_0^{(2)}$ and $E_0^{(4)}$, and by substituting the results from the constant density of states, we obtain
\begin{equation}
\begin{split}
E(\varphi) = \left(-E_J + \frac{2T^4\mathcal{D}^2}{\Delta}\left(\frac{\pi^2}{16} -\alpha\mathcal{D}^2\Delta^2     \right)\right)\cos\varphi + \\ + \frac{2T^4\mathcal{D}^2}{\Delta}\left( \alpha' -\alpha''\mathcal{D}^2 \Delta^2    \right)\cos(2\varphi) .
\end{split}
\end{equation}

}

\end{document}